\documentclass[12pt]{article}
\catcode`\@=11

\@addtoreset{equation}{section}
\def\theequation{\arabic{section}.\arabic{equation}}
     
\def\eqnarray{\let\@currentlabel=\theequation\refstepcounter{equation}
    \global\@eqnswtrue
    \global\@eqcnt\z@\tabskip\@centering\let\\=\@eqncr
    $$\halign to \displaywidth\bgroup\@eqnsel\hskip\@centering
      $\displaystyle\tabskip\z@{##}$&\global\@eqcnt\@ne 
       \hfil${{}##{}}$\hfil
      &\global\@eqcnt\tw@ $\displaystyle\tabskip\z@{##}$\hfil 
       \tabskip\@centering&\llap{##}\tabskip\z@\cr}
\def\lefteqn#1{\hbox to 4\arraycolsep{$\displaystyle #1$\hss}}
\long\def\@makefntext#1{\parindent 0cm\noindent
\hbox to 1em{\hss$^{\@thefnmark}$}#1}

\def\IR{{\hbox{{\rm I}\kern-.2em\hbox{\rm R}}}}
\def\IH{{\hbox{{\rm I}\kern-.2em\hbox{\rm H}}}}
\def\IC{{\ \hbox{{\rm \shortmid}\kern-.6em\hbox{\bf C}}}}
\def\IZ{{\hbox{{\rm Z}\kern-.4em\hbox{\rm Z}}}}
\def\IN{{\hbox{{\rm I}\kern-.2em\hbox{\rm N}}}}
\def\rref#1{(\ref{#1})}
\newcommand{\beq}{\begin{equation}}
\newcommand{\eeq}{\end{equation}}
\usepackage{amsfonts,amssymb,cite,epsfig,graphicx}
\begin{document}

%
%
%
%
\def\citen#1{%
\edef\@tempa{\@ignspaftercomma,#1, \@end, }
\edef\@tempa{\expandafter\@ignendcommas\@tempa\@end}%
\if@filesw \immediate \write \@auxout {\string \citation {\@tempa}}\fi
\@tempcntb\m@ne \let\@h@ld\relax \let\@citea\@empty
\@for \@citeb:=\@tempa\do {\@cmpresscites}%
\@h@ld}
%
\def\@ignspaftercomma#1, {\ifx\@end#1\@empty\else
   #1,\expandafter\@ignspaftercomma\fi}
\def\@ignendcommas,#1,\@end{#1}
%
%
\def\@cmpresscites{%
 \expandafter\let \expandafter\@B@citeB \csname b@\@citeb \endcsname
 \ifx\@B@citeB\relax 
    \@h@ld\@citea\@tempcntb\m@ne{\bf ?}%
    \@warning {Citation `\@citeb ' on page \thepage \space undefined}%
 \else
    \@tempcnta\@tempcntb \advance\@tempcnta\@ne
    \setbox\z@\hbox\bgroup 
    \ifnum\z@<0\@B@citeB \relax
       \egroup \@tempcntb\@B@citeB \relax
       \else \egroup \@tempcntb\m@ne \fi
    \ifnum\@tempcnta=\@tempcntb 
       \ifx\@h@ld\relax 
          \edef \@h@ld{\@citea\@B@citeB}%
       \else 
          \edef\@h@ld{\hbox{--}\penalty\@highpenalty \@B@citeB}%
       \fi
    \else   
       \@h@ld \@citea \@B@citeB \let\@h@ld\relax
 \fi\fi%
 \let\@citea\@citepunct
}
%
\def\@citepunct{,\penalty\@highpenalty\hskip.13em plus.1em minus.1em}%
%
%
\def\@citex[#1]#2{\@cite{\citen{#2}}{#1}}%
%
%
\def\@cite#1#2{\leavevmode\unskip
  \ifnum\lastpenalty=\z@ \penalty\@highpenalty \fi 
  \ [{\multiply\@highpenalty 3 #1
      \if@tempswa,\penalty\@highpenalty\ #2\fi 
    }]\spacefactor\@m}
\let\nocitecount\relax  
%
\begin{titlepage}
\vspace{.5in}
\begin{flushright}
UCD-02-4\\
March 2002\\
gr-qc/0203026\\
\end{flushright}
\vspace{.5in}
\begin{center}
{\Large\bf
 Spacetime Singularities\\[.6 ex] in (2+1)-Dimensional Quantum Gravity}\\
\vspace{.4in}
{Eric~Minassian\footnote{\it email: eminassi@landau.ucdavis.edu}\\
       {\small\it Department of Physics}\\
       {\small\it University of California}\\
       {\small\it Davis, CA 95616}\\{\small\it USA}}
\end{center}

\vspace{.5in}
\begin{center}
{\large\bf Abstract}
\end{center}
\begin{center}
\begin{minipage}{4.75in}
{\small
The effects of spacetime quantization on black hole and big bang/big crunch singularities can be studied using new tools from (2+1)-dimensional quantum gravity. I investigate effects of spacetime quantization on singularities of the (2+1)-dimensional BTZ black hole and the (2+1)-dimensional torus universe. Hosoya has considered the BTZ black hole, and using a ``quantum generalized affine parameter'' (QGAP), has shown that, for some specific paths, quantum effects ``smear'' the singularity. Using generic gaussian wave functions, I show that both BTZ black hole and the torus universe contain families of paths that still reach the singularities with a finite QGAP, suggesting that singularities persist in quantum gravity. More realistic calculations, using modular invariant wave functions of Carlip and Nelson for the torus universe, further support this conclusion.
}
\end{minipage}
\end{center}
\end{titlepage}
\addtocounter{footnote}{-1}
\section{Introduction}

The issue of spacetime singularities is arguably the most fundamental outstanding problem in general relativity. Singularities are the only known instances where general relativity fails and loses predictability. The two major achievements of twentieth century theoretical physics, quantum field theory and general relativity theory, are plagued by seemingly insurmountable problems of infinities. General relativity exhibits spacetime singularities, while quantum field theory faces incurable divergences when applied to gravity. It is often hoped that a union of these theories into a quantum theory of gravity will overcome both of these failings.

Many solutions to the Einstein field equations have been found, and wide classes of these spacetimes exhibit a variety of singular behavior \cite{he, wald, joshi, chandra, kramer}. Hawking and Penrose \cite{hp} proved that under very general and physically reasonable conditions, singularities could form, and indeed are generic feature of general relativity. That is generically, there are regions where the usual descriptions of spacetime breaks down, and the laws of physics lose their predictability. The fact that ``general relativity contains within itself the seeds of its own destruction'' \cite{earman}, is so troubling that John Wheeler has called the problem of spacetime singularities ``the greatest crisis in physics of all times'' (see chapter 44 in \cite{mtw}).

However, before reaching the singularity where the classical theory fails, we enter the microscopic regime where the laws of quantum gravity should apply. Certainly, at the Planck scale, where curvature of spacetime can become enormous and the fluctuations of spacetime may even render the smooth manifold picture of general relativity inappropriate, one would not expect the classical theory to provide reliable insight. The appropriate tools to study the nature of singularities must be found within the framework of a quantum theory of gravity. Any attempt to apply quantum gravity to the problem of singularities is a step in the right direction.

Unfortunately, as of now there is no fully-understood, consistent and finite quantum theory of gravity for (3+1)-dimensional spacetime, and applications of different quantization schemes to the problem of singularities have given inconclusive and mixed results \cite{wald, clarke, senovilla}. However, with recent advances, there are now fully consistent theories of quantum gravity in (2+1)-dimensional spacetime. In fact by one account there are at least 15 of them \cite{carlip1, carlip2}. In addition, (2+1)-dimensional solutions exhibit black hole type and cosmological type singularities \cite{carlip1}. Therefore, we have all the required ingredients to investigate the problem of spacetime singularities using full quantum theories of gravity. In addition, in 2+1 dimensions we have the fortune of having enormously simplified calculations, which in 3+1 dimensions seem quite intractable. It is surprising that there have been only a few attempts in this direction \cite{hosoya, hosoya&oda, oda, puzio}. The purpose of this paper, in addition to reporting some recent results, is to point out the possible opportunities.

In this paper (2+1)-dimensional quantum gravity is applied to study the spacetime singularities present in a black hole model and a cosmological model, where the spacetime is fully quantized. Using a quantum criterion for singularities, we investigate the resolution or persistence of spacetimes singularities in a quantum theory of gravity. It seems that singularities are not resolved in these models, at least by the given definition. Persistence of singularities is demonstrated in different spacetime models, for variety of lightlike and timelike geodesics and non-geodesics, and with different quantum states.

\section{Spacetime Singularities}

There is no single, universally accepted, and fully satisfactory definition of a spacetime singularity even in classical general relativity. Roughly speaking, a spacetime singularity is a physically accessible ``region of spacetime'' where a physically relevant quantity behaves badly, usually becoming infinite at the singular point. For example, energy density and spacetime curvature (given by invariant scalar curvature polynomials) can become unbounded at the singular points. Moreover, one should require that an observer can be in causal contact with the singularity within a finite ``time'', so that the singularity in principle can affect physical experiments, and therefore be considered physically genuine. Evolution of the definition of singularity itself, to a large degree, reflects the history of evolution of ideas in this area \cite{earman,mtw, clarke, senovilla,tipler, geroch, geroch&horowitz}. The choice of a definition is a key issue, since obviously, the question of existence of singularities depends on how one defines them. This is not a case of a circular argument, but a real problem, closely related to the question of nature of singularities, and how one can formulate a criterion that is both universal enough to cover the known examples and narrow enough to exclude the false signals \cite{wald, geroch, senovilla}.

\subsection{Definition in General Relativity}

A widely used and intuitively simple definition of a singularity in general relativity is the idea of \textit{geodesic incompleteness}. Spacetime is geodesically incomplete (\textit{g-incomplete}) if it contains at least one geodesic that is inextendible (in at least one direction) and has a finite affine length. This means that test particles or free falling observers reach the ``end of spacetime'' in a finite ``time'' (proper time for timelike geodesics) and fall off the universe or cease to exist. To exclude artificially removed points one considers only \textit{inextendible spacetimes} \cite{wald}.

Although g-incompleteness indicates a serious physical pathology, it is neither an all-inclusive definition, nor even all-exclusive of false alarms. There are examples of compact spacetimes that are g-incomplete, which have no mathematically meaningful ``holes'' present (see \cite{misner} or problem 2, chapter 9 in \cite{wald}). On the other hand, there are also examples of timelike, null, and spacelike geodesically complete spacetimes which contain future inextendible timelike accelerated curves, with bounded acceleration and finite ``time'' \cite{geroch}. That is, one can have a physically acceptable non-inertial observer, who by traveling along such a curve can reach the end of spacetime in a finite proper time, using only rockets with finite amount of fuel and acceleration. Furthermore, the concept of geodesic is a classical one. Since our goal is to apply quantum theory, it would not be suitable to start with a definition that already contains ingredients that are not naturally generalizable to quantum mechanical situations.

A more suitable definition is the idea of \textit{b-incompleteness} (bundle-incompleteness) first introduced by Schmidt \cite{schmidt} and later used as a more powerful tool to detect singularities (for example the criteria used in Hawking and Ellis (1973) \cite{he}). Here, we replace a timelike or null geodesic with a general $C^1$ causal curve. Now, to have a measure of ``time'' or length of a general curve, we need to replace affine parameter, which is applicable to geodesics, with a generalized affine parameter (GAP) that can be applied to a general curve \cite{schmidt}.The generalized affine parameter measuring the ``length'' $\left| \gamma \right|$ of a general causal curve $\gamma \left( t \right)$, where $\gamma :\left[ {0,1} \right) \to M$, is given by:
\beq
\left| \gamma  \right| = \int\limits_0^1 {dt\sqrt
{\sum\limits_{i = 0}^3 {V^i \left( t \right)V^i \left( t \right)}
} }.
\label{e2.1}
\eeq
Here, $V^i \left( t \right)$ are the components of the tangent vector to the curve $\gamma \left( t \right)$ in an orthonormal basis, with the basis (tetrad frame)carried along the curve by parallel propagation. Note that the quantity under the square root is positive definite, so $\left| \gamma \right|$ can be used to define the length of any curve; in particular, for null curves we get a nonzero measure. At first it seems that the GAP is not useful, since it is clearly not a coordinate invariant quantity, and depends on arbitrary choice of the initial basis and the initial point $t=0$. However, one can show \cite{he} that the ``length'' of a given curve $\gamma$ is finite in terms of a generalized affine parameter if and only if it is finite for any other generalized affine parameter. Therefore, finiteness or infiniteness of the GAP is an invariant concept. In addition, if the curve is a non null geodesic, finiteness or infiniteness of the generalized affine parameter agrees with that of affine parameter.

Now, we can define the concept of a missing point or a ``hole'' for a general $C^1$ curve. We say a spacetime is \textit{b-incomplete} if there is an inextendible curve that has a finite generalized affine parameter, indicating a ``hole'' that can be reached by an observer in a finite ``time''. Similarly, a spacetime is \textit{b-complete}, and defined to be \textit{singularity free}, if all incomplete curves with a finite GAP have endpoints (no missing points) \cite{he}. We can weaken this by allowing spacelike b-incomplete curves, but taking causal incomplete curves as a criterion of a ``physical'' singularity. The concept of b-completeness can be defined for any manifold M, as long as there is a connection on M.

\subsection{Definition in Quantum Theory}

To study classically singular spacetimes in quantum gravity, one needs a definition of a singularity in quantum gravity. Hosoya used the classical definition of b-incompleteness given above, and generalized it to the quantum case \cite{hosoya}.

The key idea required here is a physically meaningful measure of a ``time parameter'' or ``length'' of a curve (analogous to the generalized affine parameter) in quantum gravity, where the spacetime metric itself becomes a quantum variable and the geometry of spacetime is subject to fluctuations. If we have such a measure, then we can look for any inextendible curve that quantum mechanically still has a finite ``length'', indicating a pathology, a tell-tell signs of presence of ``holes'' in now quantized spacetime. If this is the case, we say we have a quantum b-incompleteness, and a singularity is present \cite{hosoya}.

Hosoya proposes a quantum mechanical generalization of the GAP, which we call the quantum generalized affine parameter (QGAP), or a quantum b-length $\left| \gamma  \right|_q$ given below:

\beq 
\left| \gamma  \right|_q  = \int\limits_0^a {dt\sqrt
{\sum\limits_{i,j = 0}^3 {\left\langle {\left( {V^\mu~e_\mu^j
\left( 0 \right)\left[ {P\exp \int_{}^t \omega } \right]_{\;j}^i
} \right)^2 } \right\rangle } } }
\label{e2.2}
\eeq
Here
$\left\langle {~} \right\rangle$ indicates an expectation value with respect to some quantum state of the geometry $\left| \Psi \right\rangle$, $V^\mu$ is the tangent vector to the curve at point $\gamma \left( t \right)$, $e_\mu ^j \left( 0 \right)$ some initial choice of tetrad basis at the initial point $\gamma \left( 0 \right) $, and $\left[ {P\exp \int^t \omega } \right]_{\;j}^i$ is the path-ordered integral of the spin connection $\omega _{\; j}^i = \left( {\omega _\nu} \right)_{\;j}^i dx^\nu$ (see Appendix C.1 in \cite{senovilla} for a definition). The Greek indices are spacetime indices, and Latin indices refer to the tetrad basis.

Note that the only difference between the QGAP given by \rref{e2.2}, and the GAP given by \rref{e2.1} is the expectation value. To obtain $e_\mu ^i \left( t \right)$ at any point $\gamma \left( t \right)$, we start from $e_\mu ^j \left( 0 \right)$ at the initial point $\gamma \left( 0 \right) $ and evolve it by parallel propagating along the curve, using the path-ordered integral of the spin connection as our ``evolution operator'',
\beq 
e_\mu ^i \left( t \right) = e_\mu ^j \left( 0 \right)\left[
{P\exp \int_{}^t \omega} \right]_{\;j}^i . 
\label{e2.3} 
\eeq
In quantum gravity, the tetrads and spin connections become quantum operators, making the object inside $\left\langle {~} \right\rangle$,
\beq
f = \sum\limits_{i,j = 0}^3 {\left( {V^\mu\,e_\mu ^j \left( 0 \right)\left[ {P\exp \int_{}^t \omega  } \right]_{\;j}^i } \right)^2} 
\label{e2.4} 
\eeq
an operator, which acts on a state of geometry, giving an expectation value:
\beq
\left\langle {\hat f} \right\rangle  = \left\langle \Psi
\right|\hat f\left| \Psi \right\rangle .
\label{e2.5}
\eeq
Therefore, quantum mechanically what we measure at each point $\gamma \left( t \right)$ is an expectation value of a quantum operator, which classically corresponds to the infinitesimal b-length at that point, and by integrating along the path as the observer (or a classical test particle) moves forward, we find the total b-length of the path traveled in a fully quantized spacetime. In a sense, we study motion of a classical observer in quantum geometry. 

This is a complimentary approach to other works that look at motion of a quantum particles in a classical background (for an interesting example of the latter approach see \cite{horowitz&marolf}). The present approach is more in the spirit of quantum gravity and takes advantage of the tools from (2+1)-dimensional quantum gravity, while the latter approach is motivated by string theory and known solutions corresponding to motion of quantum test strings in a classical background geometry \cite{horowitz&marolf}. Ideally, a fully quantum mechanical treatment of a quantum particle in a quantized spacetime background would be desirable. Some investigations are in progress in this direction.

\section{Quantum Gravity in 2+1 Dimensions}

There are several working theories of quantum gravity in (2+1)-dimensional spacetime (which are not all equivalent) \cite{carlip1}. It is only natural to take advantage of these theories and examine the unresolved problems of quantum gravity. However, the physical spacetime seems to be (3+1)-dimensional, and it is not obvious if such considerations in (2+1)-dimensional gravity would be relevant to physical reality. Therefore, we have to ask to what extent quantum gravity in (2+1)-dimension is useful in analysis of singularities.

In 2+1 dimensions, starting from the standard form of Einstein-Hilbert action one derives the Einstein field equations (same form for any dimension $n \ge 3$). The general relativity thus obtained is generally covariant, and its gauge group is the diffeomorphism group. Similar to (3+1)-dimensional gravity, nontrivial global and topological features and spacetime singularities can arise in 2+1 dimensions.

The main difference from (3+1)-dimensional gravity is that the Weyl tensor is identically zero in 2+1 dimensions, so the Riemann curvature tensor is completely determined in terms of Ricci tensor. Physically, this means that there are no local dynamical degrees of freedom. Hence, gravity does not propagate (no gravitational waves, or gravitons in the quantum theory), and curvature is concentrated only at the location of the source. Not surprisingly, in the Newtonian limit, test particles experience no Newtonian force.

However, these differences do not make (2+1)-dimensional gravity trivial. For example, moving particles scatter nontrivially; and spacetimes with nontrivial fundamental groups give rise to a finite number of global degrees of freedom, which make construction of highly tractable quantum theories possible. Even the local degrees of freedom can be restored by introducing a dilation or a gravitational Chern-Simons term. Therefore, many of the conceptual issues arising in a realistic (3+1)-dimensional gravity can be studied in 2+1 dimensions \cite{carlip1, brown}.

The question of resolution or persistence of spacetime singularities is of such conceptual nature, and the essential features may not depend on details of 2+1 or 3+1 theories. For example, many expect that the essential ingredients of a quantum theory, such as uncertainty principle, are incompatible with formation of singularities. Here, by analyzing specific singular spacetimes in (2+1)-dimensional quantum gravity, I show that quantization of spacetime does not seem to be sufficient to resolve some of the spacetime singularities, at least in this scheme. Such counterexamples challenges the belief that a quantum theory of gravity is irreconcilable with any spacetime singularities. Of course, there might be some feature that could still cause the quantum gravity in 3+1 dimensions to resolve all singularities, but that would be a peculiarity of the particular dimensions rather than a fundamental necessity of a quantum theory.

\section{Black Hole Type Singularities}

Hosoya has suggested that if we quantize a spacetime that is classically singular (in the sense of b-incompleteness), quantum mechanical fluctuations of the connections may become very large as we approach a region that is classically singular \cite{hosoya}. The fluctuations may become so violent that the ``length'' (the QGAP) of the curve $\gamma$ might become infinite. The QGAP becoming infinite indicates that the singularity is ``pushed away to infinity'' and a physical observer will never reach the ``singularity'' in a finite ``time.'' The ``singularity'' becomes physically inaccessible and quantum mechanically the spacetime becomes b-complete, and by definition, singularity free. Using the BTZ model of black hole in 2+1 dimensions \cite{BTZ} and a particular choice of a path, he found that this seems to be the case \cite{hosoya,hosoya&oda}.

However, this seems to be due to the particular choice of the path and is not a general feature of paths in this spacetime. In fact I show that there are paths that reach the classically singular region within a finite QGAP in quantized BTZ spacetime, rendering the spacetime quantum mechanically b-incomplete, and singular.

\subsection{The (2+1)-Dimensional BTZ Black Hole}

The spacetime metric for the BTZ black hole is given by
\begin{eqnarray}
ds^2 & = & - \frac{1}{{1 - \frac{{\hat r^2 }}{{l^2 }}}} d\hat r^2  + \alpha ^2 \left( {1 - \frac{{\hat r^2 }}{{l^2 }}} \right)d\hat t^2  + \beta ^2 \hat r^2 d\hat \theta ^2 \label{e4.1} \\
  & = & - (e^0 )^2  + (e^1 )^2  + (e^2 )^2 . \nonumber
\end{eqnarray}
Here, we consider a black hole in presence of a negative cosmological constant $\Lambda  =  -1/l^2$, with zero angular momentum and zero electric charge, where $\alpha$ and $\beta$ are adjustable parameters (constants on shell). We have, $\beta^2 = M$, where $M$ is the mass of the black hole, and $\alpha$ is roughly related to the Hawking temperature or the ``opening angle'' [26, 11]. In spirit of minisuperspace quantization, we take $\alpha$ and $\beta$ to be arbitrary parameters. The coordinates here are related to the more common ones by:
\beq
\hat \theta =\sqrt M \,\theta ,\;\hat t = \sqrt M t,\;\hat r
= r/\sqrt M .
\label{e4.2}
\eeq
Here, I follow the particular choice of coordinates and notation given in \cite{hosoya}. To avoid confusion, note that the hat on the coordinates $\left( {\hat t, \hat r, \hat \theta } \right)$ only indicate a particular scaling introduced in \cite{hosoya}, and has nothing to do with the operator symbol used below. The black hole has a spacelike causal singularity at $\hat r = 0$, and an event horizon at $\hat r = l$. The singularity arises from identification of $\hat \theta  = 0$ with $\hat \theta  = 2\pi \sqrt M $. The curvature at the singular point does not become infinite, so this is a rather mild singularity.

\subsection{Triads, Spin Connections, and the First Order Formalism}

To apply first order formalism and proceed to quantization, we need to calculate the triads and the spin connections. We can read off triads from \rref{e4.1},

\beq \left\{
\begin{array}{lll}
   e^0 & = & d\hat r/\sqrt {1 - \hat r^2 /l^2 } \\
   e^1 & = & \alpha \sqrt {1 - \hat r^2 /l^2 } dt .\\
   e^2 & = & \beta \, \hat r \, d\hat \theta
\end{array}
\right. 
\label{e4.3}
\eeq
Now, we impose the torsion-free condition for the first Cartan structure equation \cite{carlip1}, and calculate the spin connection 1-forms

\beq
\left\{ 
{\begin{array}{ll}
   \omega _{\; 1}^0 & = - \alpha \, \hat r /l^2 \, d\hat t \\
   \omega _{\; 2}^0 & = \beta \sqrt {1 - \hat r^2 /l^2 } \, d\hat \theta .\\
   \omega _{\; 2}^1 & = 0
\end{array}} 
\right.
\label{e4.4}
\eeq
Since we are interested in the region close to the classical singular point, we only consider the region $\hat r \le l$, inside the horizon. From the metric \rref{e4.1}, we can see that $\hat r$ becomes the timelike coordinate inside the horizon, and therefore we use it as the time variable in our quantization, assuming that the parameters $\alpha$ and $\beta$ are functions of $\hat r$. Now, we can use the second Cartan structure equation \cite{carlip1} to calculate the curvature two-forms
\beq
\left\{ 
{\begin{array}{ll}
   R_{\;1}^0 & = -R^2 = - \frac{1}{l^2}\left( 
   {1 + \frac{\dot \alpha \hat r}{\alpha }} \right)e^0 e^1 \\
   R_{\;2}^0 & = R^1 = - \left( 
   {\frac{1}{l^2} - \frac{\dot \beta }{\beta \hat r}
   \left( {1 - \hat r^2 /l^2 } \right)} \right)e^0 e^2 ,\\
   R_{\;2}^1 & = R^0 = - \frac{1}{l^2}e^1 e^2
\end{array}} \right.
\label{e4.5}
\eeq
where we used dot to indicate a derivative with respect to our time parameter $\hat r$.

\subsection{Quantization}

Now, we apply canonical quantization procedure. In terms of curvature two-forms and triads, the Einstein action is
\begin{eqnarray}
 S &=&- \frac{2}{16\pi G}\int \limits_M {\left( { - e_a R^a - \Lambda e^0 e^1 e^3} \right)} \nonumber\\ 
   &=& - \frac{2}{16\pi G}\int \limits_0^{2\pi} {d\theta \int\limits_0^T {dt \int\limits_0^r {dr \,\alpha \,\dot \beta } }} = - \frac{T}{4G}\int\limits_0^r {dr\,\alpha \,\dot \beta } ,\label{e4.6}
\end{eqnarray}
from which we obtain
\beq
p_\beta = \frac{\partial L}{\partial \dot \beta} = - \frac{T}{4G}\alpha .
\label{e4.7}
\eeq
Now, applying a simple canonical quantization to the Poisson bracket $\left\{ {\beta ,p_\beta} \right\} = 1\, \to \, [\hat \beta ,\hat p_\beta] = i\hbar$, we obtain the commutation relation
\beq
\left[ {\hat \beta ,\hat \alpha } \right] = i\frac{4G}{T}\hbar {\kern 1pt} .
\label{e4.8}
\eeq
Since classically our triads and spin connections contained $\alpha$ and $\beta$, canonical quantization of $\alpha$ and $\beta$ turns triads and spin connections into quantum operators
\beq
e^a ,\omega _{\; b}^a  \to \hat e^a ,\hat \omega _{\; b}^a .
\label{e4.9}
\eeq
So the function $f$ used in the definition of the QGAP given by \rref{e2.4} also becomes a function of operators, operating on a state of geometry and giving an expectation value
\beq
\left\langle {\hat f\left( r \right)} \right\rangle
 = \int {\Psi^* \left( \beta  \right) \hat f \left( {\hat \beta , \hat p_\beta ; r} \right)} \Psi \left( \beta  \right) d\beta 
\label{e4.10}
\eeq
from which we calculate the desired QGAP
\beq
\left| \gamma  \right|_q  = \int\limits_{r < l}^0 {dr' \sqrt {\sum\limits_{i,j = 0}^2 {\left\langle {\hat f \left( r' \right)} \right\rangle }}} . 
\label{e4.11}
\eeq
Note that here our time parameter is the radial coordinate $r$, and we are integrating from some value of $r$ inside the horizon, approaching the classically singular region at $r=0$, as we move along some curve $\gamma \left( r \right) \subset M$. 

\subsection{Paths and Wave Functions}

Now the final ingredients needed to calculate the QGAP and examine question of quantum b-incompleteness are a causal path and a quantum state of geometry.

A curve on our spacetime manifold is described in terms of the coordinate charts defined on the manifold, independent of the spacetime metric. Therefore, mathematically describing a curve on our manifold does not present a problem a priori in quantum geometry. However, requiring a test particle to move on a single path means that we are considering motion of a classical particle. Although the spacetime is fully quantized, our test particle is classical, and the picture is not fully quantum mechanical. However, within the context of ordinary (Copenhagen variety) quantum mechanics we are stuck with a classical observer as a fact of life. We interpret our path as path of a classical observer moving through spacetime as the geometry of spacetime fluctuates quantum mechanically. The observer makes myriads of quantum mechanical observations on the state of geometry as he moves along the path, experiencing the averaged effects of curvature of the geometry. There seem to be room for improvement for a more fully quantum mechanical picture, but I will not address that issue in this paper (see section 6).

\subsubsection{Hosoya's Accelerated Causal Path}

Hosoya used a particular path to investigate the issue of quantum b-(in)completeness of BTZ black hole. He chose the following path, that approaches the singularity at $r = 0$,
\beq
\hat t = const. \;,\quad \hat \theta _k \left( \hat r \right) = k\int_{}^{\hat r} {\frac{d\hat r'}{r' \sqrt {1 
- \frac{\hat r'^2}{l^2}}}} \;,\quad \hat r \to 0 .
\label{e4.12}
\eeq
We can see that by substituting \rref{e4.12} into the metric \rref{e4.1} we obtain
\beq
ds^2  = \frac{\left( \beta^2 k^2 - 1 \right)}{1 - \frac{\hat r^2}{l^2}}  d\hat r^2 
\label{e4.13}
\eeq
and by the following choice of parameter $k$ (inside the event horizon) we can select the path to be timelike, null, or spacelike
\beq
\left| k \beta \right| = \left\{ 
{\begin{array}{lll}
    < & 1\quad & timelike \\
    = & 1\quad & lightlike \; .\\
    > & 1\quad & spacelike \\
\end{array}} \right.
\label{e4.14}
\eeq
Now we calculating the GAP for causal curves, where we approach the singularity at $r = 0$, from a nearby point at $r = \varepsilon$, where $\varepsilon$ is small but finite. Note that all we need to know is whether the GAP is finite or infinite.
\begin{eqnarray}
  \left| \gamma  \right|_c &=& \lim \limits_{\hat r \to 0} \int_{\varepsilon} ^{\hat r} {dr'\sqrt {\sum\limits_{i,j = 0}^2 
{\left( {V^\mu \, e_\mu ^j \left( 0 \right) \left[ {P\exp \int_{}^{r'} \omega  } \right]_{\;j}^i } \right)^2 } } } \nonumber\\
&\mathop \simeq \limits_{\hat r \to 0}& \int_\varepsilon ^{\hat r} {d\hat r' \frac{\left(1 - \left| k \beta \right| \right)}{\sqrt 2} 
e^{- \left| \beta \hat \theta \left( \hat r' \right) \right|}} \;
\mathop  \simeq \limits_{\hat r \to 0} \int_\varepsilon ^{\hat r} {d\hat r' \frac{\left( {1 - \left| k \beta \right|} \right)}{\sqrt 2}
e^{ - \left| k \beta \right| \ln \hat r'}} \nonumber\\
&\mathop \simeq \limits_{\hat r \to 0}& \frac{\left( 1 - \left| k \beta \right| \right)}{\sqrt 2} \int_\varepsilon ^{\hat r} 
{d\hat r' \hat r'^{ - \left| k\beta \right|}}
\label{e4.15}
\end{eqnarray}
Therefore, we see that in case of a causal curve we get a finite value for the GAP indicating that the singularity is reached within a finite time and that a physical singularity exists.
\beq
\begin{array}{llll}
For\; causal\; curves & \left| k \beta \right| \le 1, \quad \left| \gamma  \right|_c <   \infty & \mapsto & \quad b - incomplete \\ 
For\; spacelike\; curves & \left| k\beta \right| > 1, \quad \left| \gamma  \right|_c \to \infty & \mapsto & \quad b - complete \\
\end{array}
\label{e4.16}
\eeq
To examine the quantum case, Hosoya used a gaussian wave packet, centered around a mass $M_0 =\beta _0^{{\kern 1pt}\; 2}$ with a width $\Delta$
\beq
\Psi \left( \beta \right) = \exp \left[ { - \frac{\left( \beta  - \beta _0 \right)^2}{2\Delta }} \right].
\label{e4.17}
\eeq
Now calculating the QGAP we get
\begin{eqnarray}
\left| \gamma  \right|_q &=& \mathop {\lim }\limits_{\hat r \to 0} \int_\varepsilon ^{\hat r} {dr'\sqrt {\sum\limits_{i,j = 0}^2 {\left\langle \Psi  \right| \left( V^\mu \, e_\mu ^j \left( 0 \right) \left[ P\exp \int_{}^{r'}\omega \right]_{\;j}^i \right)^2 \left| \Psi \right\rangle}} }\nonumber \\ &\approx & \int_{}^0 {dr'\exp \left[ { - \left| k \beta _0 \right|\ln \hat r + \frac{\Delta }{2}\left( k \ln \hat r \right)^2 } \right]} .\label{e4.18}
\end{eqnarray}
We see that no matter how small $\Delta$ is and what the value of $|k \beta_0 |$ is, the QGAP will diverge. Therefore, although in classical case we had a finite GAP for causal curves, the quantum case of the QGAP gives infinity for any curve, causal or not.
\beq
For\;casual\;curves: \; \left| {k \beta } \right| \le 1, \; 
\begin{array}{ll}
   \left| {\gamma} \right|_c < \infty & \mapsto \quad b - incomplete\\
   \left| {\gamma} \right|_q \to \infty & \mapsto \quad b - complete\\
\end{array}.
\label{e4.19}
\eeq
It seems that singularity is resolved and a classically singular point is now infinitely far away due to effects of black hole's mass fluctuations on the geometry of spacetime. However, we show that this effect is due to the particular accelerated path that was chosen, and there are other paths that reach the singularity in a finite ``time''.

\subsubsection{Alternative Paths: Geodesics}

The simple path
\beq
\hat t = const.\;, \quad \hat \theta  = const.\;, \quad \hat r \to 0
\label{e4.20}
\eeq
is obviously a timelike geodesic inside the event horizon, where the observer is going forward (in coordinate time given by $\hat r$) into the singularity, trying to minimize his motion in spatial directions. For this path,
\beq
ds^2 = - \left( {1 - \frac{\hat r^2}{l^2 }} \right)^{ - 1} d\hat r^2 .
\eeq
and the parameters $\alpha$ and $\beta$ which become quantum operators do not appear. Clearly, any tetrad, spin connection, or object derived from the metric on this path is independent of the quantum operators, so
\beq
\left| \gamma \right|_q = \int\limits_{r < l}^0 {dr'\sqrt {\left\langle \Psi \right|\hat f\left( r' \right) \left| \Psi \right\rangle } } = 
\int\limits_{r < l}^0 {dr' \sqrt {f \left( r' \right) \left\langle \Psi \mid \Psi \right\rangle } }= \sqrt {\left\langle \Psi \mid \Psi \right\rangle}\, \left| \gamma \right|_c\, .
\label{e4.22}
\eeq
Therefore, the QGAP is exactly equal to the GAP for all normalizable states. Since the classical GAP is finite (which can be easily verified), it follows that the QGAP is also finite for the quantum case. So for this timelike geodesic, we have both classical and quantum b-incompleteness, and in fact, geodesic incompleteness.

One might object to this choice of path by saying that for a realistic quantum test particle it is awkward to impose a condition that absolutely localizes the particle in space. Small deviations from the geodesic path, due to quantum fluctuations of the particle, would introduce the quantum operators into the expectation value again, exposing the particle to the effects of fluctuations of spacetime geometry. Accumulation of such an effect might be significant, or even contribute an infinite value. So it might be argued that suppression of the quantum operators in \rref{e4.22} is artificial.

I now show that this objection is not warranted, by examining an accelerated path that has a non-zero spacelike component, and therefore is sensitive to fluctuations of geometry. I show that quantum fluctuations do not contribute an infinite effect to the QGAP for such paths.

\subsubsection{Alternative Paths: Accelerated Causal Paths}

Here, we choose the following path
\beq
\hat \theta = const.\;, \quad \hat t_k \left( \hat r \right) = k\int_{}^{\hat r} {\frac{dr'}{1 - \frac{r'^2}{l^2}}} \;,\quad \hat r \to 0 .
\label{e4.23}
\eeq
Substituting in the metric \rref{e4.1} we get
\beq
ds^2 = \left( {\alpha ^2 k^2 - 1} \right)\left( {1 - \frac{\hat r^2}{l^2}} \right)^{ - 1} d\hat r^2 
\label{e4.24}
\eeq
and inside the event horizon we have
\beq
\left| {k\alpha } \right| = \left\{ 
{\begin{array}{ll}
  { < 1\quad timelike} \\
  { = 1\quad lightlike} \\
  { > 1\quad spacelike} \\
\end{array}}. \right.
\label{e4.25}
\eeq
Calculating the GAP we get
\begin{eqnarray}
&&\left| \gamma \right|_c = \mathop {\lim }\limits_{\hat r \to 0} \int_\varepsilon^{\hat r} {dr'\sqrt {\sum\limits_{i,j = 0}^2 {\left( {V^\mu \, e_\mu ^j \left( 0 \right)\left[ {P\exp \int_{}^{r'} \omega  } \right]_{\;j}^i } \right)^2 } } }\nonumber\\
&& \mathop \simeq \limits_{\hat r \to 0} \int_\varepsilon ^{\hat r} 
{d\hat r'} \sqrt {\frac{1}{1 -\frac{\hat r'^2}{l^2}} \left( {\left( {1 + \alpha^2 k^2 } \right) \cosh \left[ {\alpha k \ln \left( {1 - \frac{\hat r'^2}{l^2}} \right)} \right] + 2\alpha k \sinh \left[ {\alpha k \ln \left( {1 - \frac{\hat r'^2}{l^2}}\right)} \right]} \right)}\nonumber\\
&& \mathop \sim \limits_{\hat r \to 0} -\sqrt {\left( {1 + \alpha^2 k^2 } \right)} \;\varepsilon + O[\varepsilon ^3 ] .\label{e4.26}
\end{eqnarray}
We only need to consider the asymptotic limit close to singularity, and it is obvious that the integral is finite for any value of $ k \alpha $, so classically as expected, we have
\beq
For\;any\;curve\,(casual\;or\;spacelike), \quad \left| \gamma \right|_c < \infty \quad \mapsto \quad b - incomplete.
\label{e4.27}
\eeq
Now we calculate the QGAP using the same gaussian wave packet given above.
The object inside the integral is a quantum operator,
\beq
\hat f\left( {\hat \alpha ; \hat r} \right) = \frac{1}{1 -\frac{\hat r^2 }{l^2}}\left( {\left( {1 + \hat \alpha^2 k^2} \right)\cosh \!\left[ {\hat \alpha k \ln \!\left( {1 - \frac{\hat r^2}{l^2}} \right)} \right] + 2\hat \alpha k\,\sinh \!\left[ {\hat \alpha k \ln \!\left( {1 - \frac{\hat r^2}{l^2}} \right)} \right]} \right)
\label{e4.28}
\eeq
where $\hat r$, the time parameter, is of course a c-number for 2+1 canonical quantum gravity, and $\alpha$ is the quantum operator conjugate to $\beta$. Although finding the expectation value of \rref{e4.28} looks rather cumbersome, fortunately all we need is the asymptotic behavior of the integral as we approach $r\sim 0$. Also by transforming to a ``momentum'' representation $\Psi (\alpha)$, we can avoid taking derivatives. The momentum representation of the wave packet is still a gaussian. After some algebra we arrive at
\beq
\left\langle {\hat f\left( {\hat r} \right)} \right\rangle = 1 + \frac{1}{\Delta}\left( {1 - \frac{1}{2\sqrt {2\Delta } }} \right)\left( {\frac{4G\hbar k}{T}} \right)^2 + O\left[ {\hat r} \right],
\label{e4.29}
\eeq
resulting in
\begin{eqnarray}
  \left| \gamma \right|_q & = & \mathop {\lim }\limits_{\hat r \to 0} \int_\varepsilon^{\hat r} {d\hat r'\sqrt {\left\langle {\hat f\left( {\hat r'} \right)} \right\rangle } }\nonumber\\ 
& \approx & \sqrt {1 + \frac{1}{\Delta }\left( {1 - \frac{1}{2\sqrt {2\Delta } }} \right)\left( {\frac{4G\hbar k}{T}} \right)^2} \int_\varepsilon^{\hat r} {d\hat r'} \left( {1 + O\left( {\hat r'} \right)} \right)\;\mathop \propto \limits_{\hat r \to 0} \varepsilon + O\left( {\varepsilon ^2} \right). \label{e4.30}
\end{eqnarray}

Therefore, the QGAP is finite, indicating that the classically singular point is reached in a finite ``time''. We see that for both classical and quantum cases this family of paths is incomplete.

Note that in the quantum case, when $|k \alpha |$ becomes an operator and the casual structure itself is subject to quantum fluctuation, the definition \rref{e4.25} of causal paths is ambiguous. However, it is reasonable to still label paths based on their classical casual nature, viewing it as an averaged characteristic of quantum paths. A natural quantum version of \rref{e4.25}, for a given a wave function, would be to replace the $\alpha$ in $| k \alpha |$ with its expectation value $| k \langle \hat \alpha \rangle|$. Similarly, we should replace $| k \beta|$ with $| k \langle \hat \beta \rangle|$ for Hosoya's choice of paths in \rref{e4.14}.

In any case, since the above family of paths was b-incomplete for both causal and spacelike curves in the classical case, and for all values of $k$, in the quantum mechanical case, this point does not alter our conclusion.

\section{Cosmological Singularities}

The BTZ black hole has only a mild singularity, which in any case is hidden from outside observers by an event horizon. If the strong Cosmic Censorship Conjecture turned out to be valid for realistic black hole singularities, then the pathological regions of the spacetime are veiled and predictability is restored to the rest of the spacetime. In that case, persistence or resolution of black hole singularities in quantum regime might not be as
consequential.

However, a ``cosmological'' singularity such as the initial big bang singularity has no event horizon shielding it from observers. In fact, for many cosmologies, all observers in the universe are in causal contact with it through past-directed inextendible causal geodesics with finite length. Therefore, the question of resolution or persistence of such singularities in quantum regime is not just a mathematical curiosity, but could have significant consequences in cosmology, and in principal be observationally verifiable. If the universe is closed, a singularity in the future, the big crunch, will also be in the future of all observers, and its nature will determine their fate. Therefore, the problem of resolution or persistence of cosmological singularities can theoretically be even more interesting than the black hole case.

The simplest cosmological model in 2+1 spacetime dimensions that has rich enough structure to have a meaningful quantum theory of gravity is a universe with spatial topology of a torus. Models with higher genus have been studied, but not much is known about their quantized states.

Once again, we shall see that the singularity is preserved in quantum regime, and that there are large classes of families of causal paths, both lightlike and timelike, geodesic and accelerated, that reach the singularity in a finite ``time.'' The spacetime structure of the (2+1)-dimensional torus universe is quite different from the BTZ black hole, both classically and quantum mechanically. Therefore, persistence of singularities in this cosmological model provides further evidence for generality of the results obtained in the black hole case.

\subsection{The (2+1)-Dimensional Torus Universe}

Consider a spacetime manifold with the topology
\beq
M = \left[ {0,1} \right] \times T^2
\label{e5.1}
\eeq
where the manifold is foliated by surfaces of constant mean extrinsic curvature. Each slice is a spacelike torus, representing a spatially closed universe. Area and modulus varies from slice to slice. We define a ``global'' time coordinate $T$ on $M$, by the York time-slicing,
\beq
T =  - Tr{\kern 1pt} K
\label{e5.2}
\eeq
where $k$ is the mean extrinsic curvature of the spatial slices. The general solution of the Einstein field equations can be obtained through ADM formalism \cite{ADM, carlip1, carlip&nelson1}, and the  metric is found to be
\beq
ds^2 = - F^4 dt^2 + F^2 e^t \left( {a dx + b dy} \right)^2 + F^2 e^{-t} \left( {\lambda dx + \mu dy} \right)^2 
\label{e5.3}
\eeq
where
\beq
F^2 = \frac{1}{{e^t  - \Lambda e^{ - t} }} = \left( {T^2  - 4\Lambda } \right)^{ - \frac{1}{2}} 
\label{e5.4}
\eeq
\beq
T = e^t  + \Lambda e^{ - t} 
\label{e5.5}
\eeq
Here, $a$, $b$, $\lambda$, and $\mu$ are constants that uniquely determine the classical geometry. We can read off triads from the metric \rref{e5.3}
\beq
\left\{ {\begin{array}{*{20}ll}
   e^0 & = F^2 dt \\
   e^1 & = Fe^{t/2} (a dx + b dy) \\
   e^2 & = Fe^{ - t/2} \left( {\lambda dx + \mu dy} \right) \\
\end{array}}, \right.
\label{e5.6}
\eeq
and from the torsion-free condition for the first Cartan structure equation we obtain the spin connections
\beq
\left\{ {\begin{array}{*{20}l}
   {\omega ^0  = 0\quad \quad }  \\
   {\omega ^1  =  - e^t e^2 \;\,}  \\
   {\omega ^2  = \Lambda e^{ - t} e^1 }  \\
\end{array}} \right..\quad \quad \quad \quad\quad\quad\!
\label{e5.7}
\eeq
Canonical quantization gives
\begin{eqnarray}
 \left[ \hat \mu ,\hat a \right] &=& - \frac{i}{2} \;\; \to \;\;\; \hat \mu ,\; \hat a = \frac{i}{2} \frac{\partial}{\partial \mu} \nonumber \\ 
 \left[ \hat \lambda ,\hat b \right] &=&\frac{i}{2} \;\;\;\;\; \to \;\; \hat \lambda ,\;\hat b = - \frac{i}{2} \frac{\partial }{\partial \lambda} \label{e5.8}
\end{eqnarray}
and we find the expectation value by
\beq
\left\langle {\hat f\left( {\hat e^a ,\hat \omega _{\; c}^b } \right)} \right\rangle = \int_{-\infty}^\infty {d\mu \int_{-\infty}^\infty {d\lambda \Psi ^* \left( {\mu ,\lambda } \right)\hat f \, \Psi \left( {\mu ,\lambda } \right)}}
\label{e5.9}
\eeq

\subsubsection{Paths and Gaussian Wave Functions}
Given a path $\gamma (t)=(x(t),y(t))$, we can evaluate $V^\mu$ and $\left[ {P\exp \int_{}^t \omega } \right]_{\;b}^a$ and calculate the GAP. To calculate the QGAP
\beq
\left| \gamma \right|_q = \int^{\infty} \! {dt\sqrt {\sum\limits_{a = 0}^2 {\left\langle {\left( {V^\mu \, e_\mu ^b \left( 0 \right) \left[ {P\exp \int_{}^t \omega  } \right]_{\;b}^a } \right)^2 } \right\rangle } } } 
\label{e5.10}
\eeq
we also need to specify a quantum state representing the state of geometry. Here again as a generic state, we use a gaussian wave packet
\beq
\Psi \left( \mu ,\lambda \right) = N\exp \left[- \frac{\left( \mu - \mu _0 \right)^2 + \left( \lambda - \lambda _0 \right)^2}{2\Delta} \right]
\label{e5.11}
\eeq
centered around some geometric state of the torus given by constants $\mu_0$ and $\lambda_0$ and with a width $\Delta$. To examine our criteria for existence of singularities, the choice of paths is critical. In addition, we choose paths that will keep the calculations tractable.

\subsubsection{Spacetime Geodesics: $\mathbf{g_0}$ Type Paths}

We can consider geodesics by choosing the path
\beq
x = const.\, ,\quad y = const.\, ,\quad t \to \infty .
\label{e5.12}
\eeq
Here in the limit $t \to \infty$ we approach the big crunch singularity. Since analysis of big bang singularity in the limit $t \to -\infty$ is the time reversal of, and mathematically identical to, the big crunch case we choose this limit without loss of generality. The metric along the curve reduces to:
\beq
ds^2  =  - F^4 dt^2 
\label{e5.13}
\eeq
Classically we have:
\begin{eqnarray}
  \left| \gamma  \right|_c &=& \mathop {\lim }\limits_{t \to \infty } \int_{}^t {dt'\sqrt {\sum\limits_{i,j = 0}^2 {\left( {V^\mu \, e_\mu ^j \left( 0 \right)\left[ {P\exp \int_{}^{t'} \omega  } \right]_{\;j}^i } \right)^2 } } } \nonumber\\
   & \mathop = \limits_{t \to \infty }& \int_{}^t {dt'} \sqrt {F^4 } \mathop = \limits_{t \to \infty } \int_{}^t {dt'(e^{t'}- \Lambda e^{-t'} )^{-1}} \mathop \sim \limits_{t \to \infty } \int_{}^t {dt' e^{ - t'}} < \infty \label{e5.14}
\end{eqnarray}
obtaining a finite GAP indicating b-incompleteness (and g-incompleteness) as expected. Obviously in the quantum case, the QGAP for geodesics reduces trivially to the classical GAP
\beq
\left| \gamma \right|_q \mathop = \limits_{t \to \infty } \int_{}^t {dt'\sqrt {\left\langle \Psi \right| \hat f\left( t' \right)\left| \Psi  \right\rangle}} \mathop = \limits_{t \to \infty } \int_{}^t {dt'\sqrt {f\left( t' \right)\left\langle \Psi | \Psi \right\rangle } } 
\label{e5.15}
\eeq
for any normalizable wave function, including Gaussian wave functions, since the operator dependence in $\hat f (t')$ has been suppressed by the choice of the path, just as in the black hole case. Therefore, in the quantum case, spacetime remains g-incomplete for these geodesics, and the singular nature of the spacetime is preserved.

\subsubsection{Accelerated Paths: Spacelike Geodesics of the Torus}

To take the above result more seriously, we need to examine the accumulative effects of spacetime fluctuations on the QGAP to see if the choice of a geodesic path is reasonable and is not likely to ignore infinite contributions. To consider paths that do not suppress the quantum operators and so are sensitive to fluctuations of spacetime, we need to consider accelerated paths in spacetime.

\begin{figure}
\begin{center}
\includegraphics{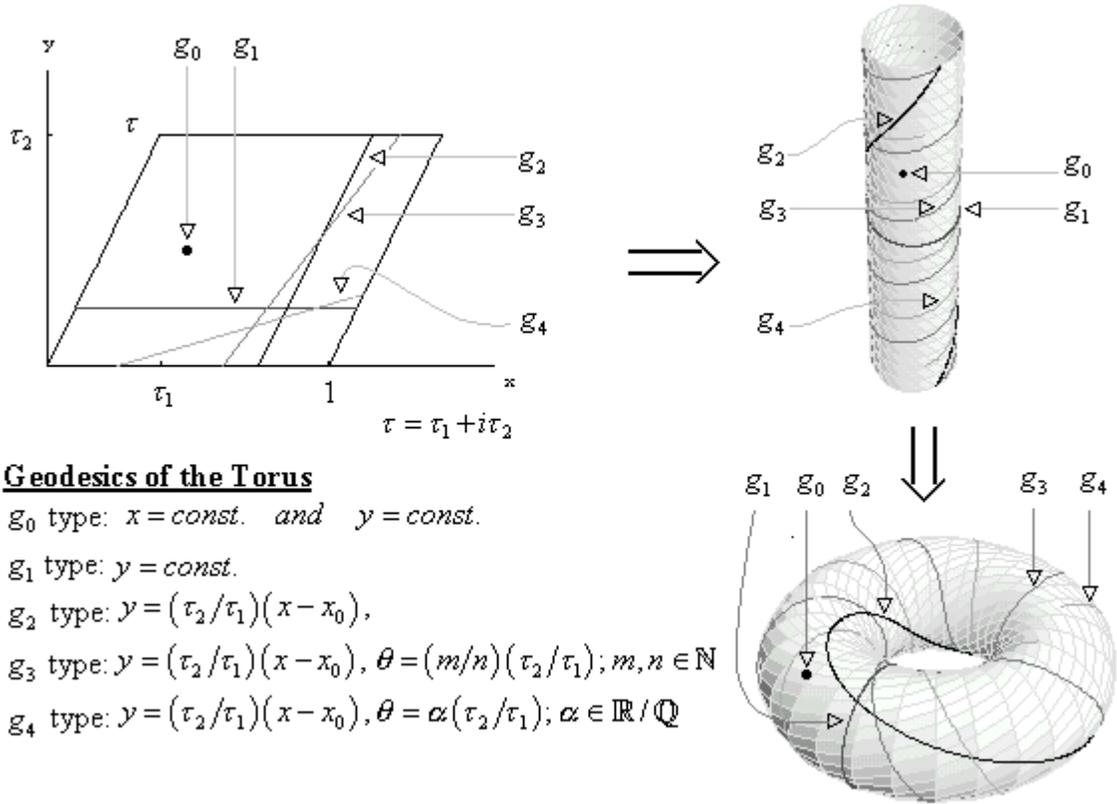}
\caption{The spatial geodesics in the (2+1)-dimensional torus universe.}
\label{torus-geodesics}
\end{center}
\end{figure}

There are two generators of the fundamental group on a torus, giving rise to four types of spatial geodesics. To keep the calculations manageable, we choose causal paths whose projections on the spacelike tori follow these geodesics, hence giving rise to four classes of causal paths. The first class ($g_1$) has as its projections the set of closed geodesic that circle once around the torus ``passing \textit{through} the hole in the middle''. The second class ($g_2$) has closed projections that ``circle once \textit{around} the hole in the middle''. The third and forth classes have spatial geodesic projections that \textit{spiral around} the torus several times before closing ($g_3$), or open spirals that never close ($g_4$) (See figure \ref{torus-geodesics}).

\subsubsection*{5.1.3.1 Accelerated Paths: $\mathbf{g_1}$ Type Paths}

Each spacelike geodesic element $g_1 [x_0 , y_0]$ of the $g_1$ class of paths defines a family $g_1 [x_0 , y_0 ; k]$ of infinitely many causal curves,
\beq
y = const.\, ,\quad x_k \left( t \right) = k\int_{}^t {\frac{F\left( t \right) dt}{\sqrt {e^t a^2  + e^{-t} \lambda ^2 } }} 
\label{e5.16}
\eeq
parameterized by a ``speed'' parameter $k$ which determines their causal nature
\beq
\left| k \right| = \left \{ 
\begin{array}{l}
   { < 1\quad timelike}  \\
   { = 1\quad lightlike} \\
   { > 1\quad spacelike} \\
\end{array} 
\right. .
\label{e5.17}
\eeq
Since in general, we have two pairs of conjugate operators here, unlike in the black hole case with a single conjugate pair, calculations and operator orderings are more involved. However, the pair of operators that appear for $g_1$ type paths are commuting, which simplifies the calculations in this section. In addition, since we only need to verify if the integral is finite or infinite, all we need to know is the asymptotic behavior of our quantities. It is easy to verify that
\beq
V^\mu = (1,\frac{k F\left( t \right) }{\sqrt{e^t a^2 + e^{-t} \lambda ^2} }, 0) \mathop \sim \limits_{t \to \infty } (1, k e^{-t} \frac{1}{a},0),
\label{e5.18}
\eeq
and rewriting the triads \rref{e5.6} as
\begin{eqnarray}
   e_0 ^b & = & (F^2 ,0,0)^b \nonumber\\ 
   e_1 ^b & = & (0,F e^{t/2} a,F e^{-t/2} \lambda )^b , \label{e5.19}\\ 
   e_2 ^b & = & (0,F e^{t/2} b,F e^{-t/2} \mu )^b \nonumber
\end{eqnarray}
gives
\beq
V^\mu \, e_\mu ^b = \left( {F^2 ,\frac{k F^2 a }{\sqrt{a^2 + e^{-2t} \lambda ^2 }}, \frac{k F^2 a}{\sqrt {e^{2t} a^2 + \lambda ^2 }}} \right)^b \mathop  \sim \limits_{t \to \infty } \left( {e^{-t} , k e^{-t} , k e^{-2t} \frac{\lambda }{a}} \right)^b .
\label{e5.20}
\eeq
The spin connection ( only $\omega _1$ is needed here) becomes
\beq
{\omega_1}_{\;\;b}^a = F\left( t \right)\left( 
{\begin{array}{*{20}c}
   0 & {\Lambda e^{-t/2} a} & {e^{t/2} \lambda } \\
   {\Lambda e^{-t/2} a} & 0 & 0 \\
   {e^{t/2} \lambda } & 0 & 0 \\
\end{array}} \right) \mathop \sim \limits_{t \to \infty } \lambda \left( {\begin{array}{*{20}c}
   0 & 0 & 1  \\
   0 & 0 & 0  \\
   1 & 0 & 0  \\
\end{array}} \right).
\label{e5.21}
\eeq
giving
\beq
\left[ {P\exp \int_{}^t \omega } \right]_{\;b}^a \sim \left( {\begin{array}{*{20}c}
   {\cosh\left( {\lambda \, x_k \left( {a, \lambda ;t} \right)} \right)} & 0 & {\sinh\left( {\lambda \, x_k \left( {a, \lambda ;t} \right)} \right)} \\
   0 & 1 & 0  \\
   {\sinh\left( {\lambda \, x_k \left( {a, \lambda ;t} \right)} \right)} & 0 & {\cosh\left( {\lambda \, x_k \left( {a, \lambda ;t} \right)} \right)} \\
\end{array}} \right)_{\;\;b}^a \;,
\label{e5.22}
\eeq
where $x_k \left( {a, \lambda ;t} \right)$ given by \rref{e5.16} becomes
\beq
x_k \left( {a,\lambda ;t} \right) \mathop \simeq \limits_{t \to \infty } k \frac{1}{a}\int_{t_0 }^t {e^{-t'} dt'} \mathop \sim \limits_{t \to \infty } k\frac{1}{a}\left( {e^{-t_0} -e^{-t} } \right) .
\label{e5.23}
\eeq
Note that even in the exact expressions in \rref{e5.20}-\rref{e5.22}, only $a$ and $\lambda$, which later become commuting operators, appear. Therefore, for this class of paths there is no issue of operator ordering.

Now after some algebra we arrive at:
\begin{eqnarray}
  f\left( t \right) & = & \sum\limits_{i,j = 0}^2 {\left( {V^\mu \, e_\mu ^j \left( 0 \right)\left[ {P\exp \int_{}^t \omega } \right]_{\;j}^i } \right)^2 } \nonumber \\ 
   & \sim & e^{-2t} \left( {\cosh \left[ {2k\left( {\frac{\lambda }{a}} \right)\left( {e^{-t_0 }-e^{-t} } \right)} \right] + k^2 } \right) + O\left[ {e^{-3t} } \right]\label{e5.24}.
\end{eqnarray}
Calculating the classical GAP results in:
\beq
\left| \gamma \right|_c = \mathop {\lim }\limits_{t \to \infty } \int_{}^t {dt'\sqrt {f\left( t' \right)} } \sim \mathop {\lim }\limits_{t \to \infty } \sqrt {\left( {\cosh\left[ {2k\left( {\frac{\lambda }{a}} \right) e^{-t_0} } \right] + k^2 } \right)} \int_{}^t {dt'e^{-t'} } < \infty .
\label{e5.25}
\eeq
Here $t_0$ is some arbitrary finite constant, indicating the initial time when we started our journey towards the singularity. As expected, classically the GAP is finite and we have a singularity at $t \to \infty$.

Now when we quantize the spacetime, the function $f(t)$ from \rref{e5.24} becomes a quantum operator
\beq
\hat f\left( {\hat \lambda ,\hat a; t} \right) \sim e^{-2t} \left( {\cosh\left[{2k\left( {e^{-t_0}-e^{-t} } \right)\left( \frac{\hat \lambda }{\hat a} \right)} \right] + k^2 } \right) + O\left[ e^{-3t} \right],
\label{e5.26}
\eeq
operating on a gaussian $\Psi (\mu, \lambda)$. In the representation \rref{e5.8}, $\hat a = \frac{i}{2} \frac{\partial}{\partial \mu}$ would appear in the denominator, but by going instead to a ``momentum'' representation
\begin{eqnarray}
 \left[ \hat a  ,\hat \mu \right] = \frac{i}{2} \quad\; &\to& \;\;\; \hat a ,\; \hat \mu = -\frac{i}{2} \frac{\partial}{\partial a} \nonumber \\ 
 \left[ \hat b ,\hat \lambda \right] = -\frac{i}{2} \;\;\, &\to& \quad \hat b ,\;\hat \lambda = \frac{i}{2} \frac{\partial }{\partial b} 
\label{e5.27}
\end{eqnarray}
we obtain
\beq
\left( {\frac{\hat \lambda}{\hat a}} \right) = \frac{i}{2} \frac{1}{a} \frac{\partial }{\partial b},
\label{e5.28}
\eeq
where the operators are commuting and ordering is unimportant. Now using a momentum space gaussian wave function, we find the expectation value:
\begin{eqnarray}
\left\langle {\hat f\left( {\hat \lambda ,\hat a; t} \right)} \right\rangle  &=& \int_{-\infty}^\infty {da\int_{-\infty }^\infty {db \; \tilde \Psi ^{*} \left( {a,b} \right) \hat f \left( {\hat \lambda , \hat a; t} \right)\, \tilde \Psi \left( {a,b} \right)}} \label{e5.29}\\ 
&\simeq& e^{-2t} \int_{-\infty}^\infty {da \int_{-\infty }^\infty {db \; \tilde \Psi^* \left( {a,b} \right)\, \left( {\cosh\left[ {2k\left( {e^{-t_0} - e^{ -t} } \right)\left( \frac{\hat \lambda }{\hat a} \right)} \right] + k^2 } \right) \tilde \Psi \left( {a,b} \right)}} \nonumber\\
&\simeq& k^2 e^{-2t} + e^{-2t} \int_{-\infty }^\infty {da \int_{-\infty}^\infty {db \; \tilde \Psi ^* \left( {a,b} \right)\, \cosh\left[ {2k\left( {e^{-t_0 } - e^{-t} } \right)\frac{i}{2}\frac{1}{a}\frac{\partial }{\partial b}} \right]\tilde \Psi \left( {a,b} \right)}}\nonumber
\end{eqnarray}
Note that the derivative of momentum gaussian brings down a factor of $b$ inside the expectation value integral. The time dependent part, which will determine finiteness or infiniteness of  the QGAP, has a leading term $e^{-2t}$ plus terms of order $e^{-3t}$ or higher. The final integral becomes:
\begin{eqnarray}
&&\left| \gamma \right|_q = \mathop {\lim }\limits_{t \to \infty } \int_{}^t {dt' \sqrt {\left\langle {\hat f\left( {\hat \lambda ,\hat a; t'} \right)} \right\rangle } } \label{e5.30}\\ 
&&\sim \mathop {\lim }\limits_{t \to \infty} \int_{}^t {dt' e^{-t'} \!\!\left( {\sqrt {k^2 + \int_{-\infty }^\infty {da \int_{-\infty }^\infty {db \; \tilde \Psi ^ * \left( {a,b} \right)\, \cos\left[{\left( {e^{-t_0 } - e^{-t'} } \right) \frac{k}{a} \frac{\partial}{\partial b}} \right] \tilde \Psi \left( {a,b} \right)} } } } + O\left[ e^{-t'/2} \right] \right) } \nonumber
\end{eqnarray}
which is clearly finite as far as its time dependent behavior is concerned, as we see from the leading $t$ dependence of the integral in the asymptotic limit when $t \to \infty$. Therefore we have a finite QGAP, indicating b-incompleteness and persistence of the singularity in the quantum regime.

\subsubsection*{5.1.3.2 Accelerated Paths: $\mathbf{g_2}$, $\mathbf{g_3}$ and $\mathbf{g_4}$ Type Paths}

The $g_2$, $g_3$, and $g_4$ families of paths (see figure 1) are given by
\beq
y = y_0 + \theta \left( {x - x_0 } \right),\quad x_k \left( t \right) = k \int_{}^t {\frac{F\left( t \right)dt}{\sqrt{e^t \left( {a + \theta \, b} \right)^2 + e^{-t} \left( {\lambda + \theta \, \mu } \right)^2 } }} 
\label{e5.31}
\eeq
with
\[
\left| k \right| = \left\{ 
\begin{array}{*{20}l}
   { < 1\quad timelike}  \\
   { = 1\quad lightlike} \\
   { > 1\quad spacelike} \\
\end{array}
\right. .
\]
Here, the choice $\theta = \frac{\tau_2}{\tau_1}$ gives $g_2$, the choice $\theta = \frac{m}{n} \cdot \frac{\tau_2}{\tau_1}$ with nonzero $m,n \in \mathbb{N}$ gives $g_3$, and $\theta = \alpha \cdot \frac{\tau _2}{\tau _1}$ where $\alpha$ is irrational gives $g_4$. Fortunately, we don't need to consider each case individually, since using an arbitrary $\theta \in \mathbb{R}$ will cover all the possible cases (even the case for $g_1$ paths where $\theta =0$). 

For the $g_1$ case in the previous section, we had only commuting operator pairs $\hat a$ and $\hat \lambda$ appearing. However, here we have both pairs of conjugate operators appearing. The issue of operator ordering is more involved, and we need to proceed more cautiously from the beginning. Again, since we only need to verify if the integrals for the GAP and the QGAP are finite or infinite, knowing the asymptotic behavior near the singularity is sufficient. This simplifies the calculations in general, and the issue of operator ordering in particular when we consider the quantum case. 

Now we find the components of the tangent vector are
\begin{eqnarray}
V^\mu & = & (1, \frac{k F \left( t \right)}{\sqrt {e^t \left( {a + \theta b} \right)^2 + e^{-t} \left( {\lambda + \theta \mu } \right)^2 } }, \frac{\theta \, k F \left( t \right)}{\sqrt {e^t \left( {a + \theta b} \right)^2 + e^{-t} \left( {\lambda + \theta \mu } \right)^2 } }) \nonumber\\ & \mathop \sim \limits_{t \to \infty } & (1, \; k e^{-t} \frac{1}{a + \theta b}, \; \theta \, k \, e^{-t} \frac{1}{a + \theta \, b}). \label{e5.32}
\end{eqnarray}
Note that $a$, $b$, and $\theta$ are independent of $t$ for the classical case, and the sign of $a + \theta b$ can be absorbed into $k$ in the numerator.

The triads are the same as in \rref{e5.19}, but now we need all the spin connections
\beq
{\omega _0}^a _{\;\; b} = 0, \quad {\omega _1}^a _{\;\; b} \mathop \sim \limits_{t \to \infty } \lambda \left( 
{\begin{array}{*{20}c}
   0 & 0 & 1  \\
   0 & 0 & 0  \\
   1 & 0 & 0  \\
\end{array}} \right),
\quad {\omega _2} ^a _{\;\; b} \mathop \sim \limits_{t \to \infty } \mu \left( {\begin{array}{*{20}c}
   0 & 0 & 1  \\
   0 & 0 & 0  \\
   1 & 0 & 0  \\
\end{array}} \right).
\label{e5.33}
\eeq
Now using the above relations we obtain,
\beq
V^\mu e_\mu ^b \; \mathop \sim \limits_{t \to \infty } \; \left( {e^{-t}, \; k e^{-t} , \;k \frac{\left( {\lambda + \theta \mu } \right)}{\left( {a + \theta b} \right)} e^{-2t} } \right)^b ,
\label{e5.34}
\eeq
and
\begin{eqnarray}
  \left[ {P\exp \int_{}^t \omega} \right]_{\;\; b}^a & = & \left[ {P\exp \int_{}^t {\left( {\omega _1 + \theta \omega _2 } \right) \frac{dx}{dt}dt} } \right]_{\;\; b}^a \nonumber\\ & \sim & \left( 
{\begin{array}{*{20}cc}{\cosh\left[ {\left( {\lambda + \theta \mu } \right) \left( {x_k - x_0 } \right)} \right]} & 0 & {\sinh\left[ {\left( {\lambda + \theta \mu } \right) \left( {x_k - x_0 } \right)} \right]} \\ 0 & 1 & 0 \\ {\sinh\left[ {\left( {\lambda + \theta \mu } \right)\left( {x_k - x_0 } \right)} \right]} & 0 & {\cosh\left[ {\left( {\lambda + \theta \mu } \right)\left( {x_k - x_0 } \right)} \right]} \\
\end{array}} \right)_{\; b}^a\;\; \label{e5.35}
\end{eqnarray}
where asymptotically, $x_k (a, b, \mu, \lambda; \;t)$ given by \rref{e5.31} above becomes
\beq
x_k \left( a,b,\mu ,\lambda ; \;t \right) \mathop \sim \limits_{t \to \infty } k\frac{1}{\left( a + \theta b \right)}\left( e^{-t_0 } - e^{- t} \right).
\label{e5.36}
\eeq
Note that in the exact expression in \rref{e5.33} and \rref{e5.34} above, all the parameters $a$, $b$, $\mu$, and $\lambda$ appear together, which will become non-commuting operators when we consider the quantum case. After some algebra we obtain:
\begin{eqnarray}
  f\left( t \right) & = & \sum\limits_{i,j = 0}^2 {\left( {V^\mu \, e_\mu ^j \left( 0 \right)\left[ {P\exp \int_{}^t \omega} \right]_{\;j}^i } \right)^2 }\nonumber\\ 
& \sim & e^{-2t} \left( {\cosh\left[ {2k\left( {\frac{\lambda + \theta \mu }{a + \theta b}} \right)\left( e^{-t_0}- e^{-t} \right)} \right] + k^2 } \right) + O\left[ e^{-3t} \right].
\label{e5.37}
\end{eqnarray}
Classically $\left( {\frac{\lambda + \theta \mu }{a + \theta b}} \right)$ is just a constant number, and calculating the classical GAP is essentially identical to the previous section. We obtain:
\begin{eqnarray}
\left| \gamma \right|_c & = & \mathop {\lim }\limits_{t \to \infty } \int_{}^t {dt'\sqrt {f\left( t' \right)} } = \mathop {\lim }\limits_{t \to \infty } \int_{}^t {dt' e^{-t'} \sqrt {\left( {\cosh\left[ {2k\left( {\frac{\lambda + \theta \mu }{a + \theta b}} \right)\left( {e^{-t_0 } - e^{-t'} } \right)} \right] + k^2 } \right)} } \nonumber \\
& \sim & \mathop {\lim }\limits_{t \to \infty } \sqrt{\left( {\cosh\left[ {2k\left( {\frac{\lambda + \theta \mu }{a + \theta b}} \right)e^{-t_0 } } \right] + k^2 } \right)} \int_{}^t {dt'e^{-t'} } < \infty . \label{e5.38}
\end{eqnarray}
As expected, classically the GAP is finite and we have a singularity at $t \to \infty$.

Now, we need to calculate the QGAP. Quantizing, the function $f(t)$ from \rref{e5.37} becomes:
\beq
\hat f\left( {\hat \mu ,\hat \lambda ,\hat a,\hat b; \;t} \right) \sim e^{- 2t} \left( {\cosh\left[ {2k\left( {e^{-t_0} - e^{-t} } \right) \left( {\frac{\hat \lambda + \theta \hat \mu}{\hat a + \theta \hat b}} \right)} \right] + k^2 } \right) + O\left[ {e^{-3t} } \right].
\label{e5.39}
\eeq
At this point, it seems that we have two problems with the expression $\left( {\frac{\hat \lambda + \theta \hat \mu}{\hat a + \theta \hat b}} \right)$. The numerator and the denominator contain operators that do not commute with each other, and the denominator operators are derivative operators. However, we can easily verify that the numerator commutes with the denominator, so there is no ambiguity in expressing the operators as in equation \rref{e5.39}. The problem of derivatives in the denominator is solved as in the previous section, by using momentum representation. We can express \rref{e5.39} as:
\beq
\hat f\left( {\hat \mu ,\hat \lambda , \hat a,\hat b; \;t} \right) \sim e^{ -2t} \left( {\cosh\left[ {2k\left( {e^{-t_0}-e^{-t}} \right)\left( {\frac{1}{\hat a + \theta \hat b}} \right)\frac{i}{2}\left( {\frac{\partial }{\partial b}-\theta \frac{\partial}{\partial a}} \right)} \right] + k^2 } \right) + O\left[ {e^{-3t} } \right].
\label{e5.40}
\eeq
When this function operates on a gaussian momentum space wave function, the expectation value becomes:
\begin{eqnarray}
&&\left\langle {\hat f\left( {\hat \mu , \hat \lambda ,\hat a,\hat b; \;t} \right)} \right\rangle = \int_{-\infty }^\infty {da\int_{-\infty}^\infty  {db \tilde \Psi ^* \left( a,b \right)\hat f \left( {\hat \mu ,\hat \lambda ,\hat a,\hat b; \;t} \right) \, \tilde \Psi \left( a,b \right)}} \nonumber\\ && \simeq e^{-2t} \int_{-\infty}^\infty {da\int_{-\infty }^\infty {db \tilde \Psi ^ * \left( a,b \right)\, \cos\left[ {k\left( {e^{ -t_0}- e^{-t}} \right)\left( {\frac{1}{\hat a + \theta \hat b}} \right)\left( {\frac{\partial}{\partial b} - \theta \frac{\partial }{\partial a}} \right)} \right]\tilde \Psi \left( a,b \right)}}+ \nonumber \\
&& \;\;\;k^2 e^{-2t} + O\left[ e^{-3t} \right]. 
\label{e5.41}
\end{eqnarray}
The key observation is that the time dependent part of the integrand, which will determine finiteness or infiniteness of  the QGAP, again has a leading term $e^{-2t}$ plus terms of order $e^{-3t}$ or higher. The final integral becomes:
\begin{eqnarray}
&&\left| \gamma \right|_q = \mathop {\lim }\limits_{t \to \infty }\int_{}^t { dt'\sqrt {\left\langle {\hat f\left( {\hat \mu ,\hat \lambda ,\hat a,\hat b; \;t'} \right)} \right\rangle } } \label{e5.42} \\ 
&&\sim \mathop {\lim }\limits_{t \to \infty } \int_{}^t \! dt'e^{-t'} \!\! \left( \sqrt {k^2 \! + \! \mathop {\int}\limits_{-\infty }^{\infty} \! {da \! \mathop {\int}\limits_{-\infty }^{\infty} {\! db \left| {\tilde \Psi \left( a,b \right)} \right|^2 \cos\left[ {k(e^{-t_0}-e^{-t'}) \frac{\theta \left( a \!- \!a_0 \right)\! - \!\left( b \!-\! b_0 \right)}{\tilde \Delta}} \right]}}} + O\left[ e^{-t'/2} \right] \right) \nonumber
\end{eqnarray}
which is again finite as far as its time dependent behavior is concerned, as we can see from the leading $t$ dependence of the integral in the asymptotic limit when $t \to \infty$. Therefore we obtain a finite QGAP, which indicates b-incompleteness in both classical and quantum case and shows that the singularity persist in the quantum regime, for all paths considered above. That is at least for gaussian wave functions we have:
\beq
\begin{array}{*{20}c}
{\left| \gamma \right|_c < \infty \quad \mapsto \quad b - incomplete}\\
{\left| \gamma \right|_q < \infty \quad \mapsto \quad b - incomplete}\\
\end{array}.
\label{e5.43}
\eeq

\subsection{Wave Functions of The (2+1)-Dimensional Torus Universe}

One might suspect that the above finite results for the QGAP may be only due to the particular gaussian wave function that we used in our calculations. In this section we use specific wave function solutions for the (2+1)-dimensional torus universe, and verify that the persistence of the spacetime singularity in the quantum regime is independent of these choices.

\subsubsection{Modular Invariant Wave Function}

Instead of a gaussian we can use the modular invariant wave functions of Carlip and Nelson \cite{carlip1,carlip3,carlip&nelson2}
\beq
\Psi \left( {\lambda ,\mu } \right) = \int\limits_{\mathcal{F}}^{} {\frac{d^2 \tau }{\tau _2 ^2}} K\left( {T; \tau , \bar \tau ; \lambda ,\mu } \right) \, \bar{\tilde{\psi}} \left( {\tau ,\bar \tau ,T} \right)
\label{e5.44}
\eeq
where
\beq
K\left( {T; \tau , \bar \tau ; \lambda , \mu } \right) = \frac{\mu - \tau \lambda}{\pi \tau _2 ^{1/2} T} \exp \left[ {-\frac{i}{\tau _2 T} \left| {\mu - \tau \lambda } \right|^2 } \right],
\label{e5.45}
\eeq
\beq
\tilde{\psi} \left( {\tau ,\bar \tau ,T} \right)= \sum\limits_\nu C_\nu e^{- i \lambda _{\nu}^{1/2} t}\; \tilde{\psi} _\nu \left( {\tau , \bar \tau } \right),
\label{e5.46}
\eeq
and $\tilde{\psi} _\nu \left( {\tau , \bar \tau } \right)$ are the automorphic modular forms of weight $-1/2$ on the torus, with eigenvalues $\lambda _\nu$ with respect to $\Delta _{-1/2}$, the Maass Laplacian of weight $-1/2$ (see \cite{carlip1} for definitions):
\beq
\Delta _{-1/2}\; \tilde{\psi} _\nu \left( {\tau , \bar \tau } \right) = \lambda _\nu \;\tilde{\psi} _\nu \left( {\tau , \bar \tau } \right).
\label{e5.47}
\eeq
We also have
\beq
\int {d\lambda \, d\mu \, \bar K} \left( {T; \tau ', \bar \tau '; \lambda , \mu} \right) K\left( {T; \tau , \bar \tau ; \lambda , \mu } \right) = {\tau _2}^2 \, \delta ^2 \left( {\tau - \tau '} \right)
\label{e5.48}
\eeq
\beq
\left\langle {\Psi \, | \; \Psi } \right\rangle = \int\limits_F^{} {\frac{d^2 \tau }{{\tau _2}^2} \left| {\tilde \psi \left( {\tau ,\bar \tau ,T} \right)} \right|^2 } = 1
\label{e5.49}
\eeq
where
\beq
\left\langle {\Psi \left| {\hat f} \right|\Psi } \right\rangle = \int_{- \infty }^\infty {d\mu \int_{-\infty}^\infty {d\lambda \; \Psi ^ * \left( {\mu , \lambda } \right) \hat f \; \Psi \left( {\mu ,\lambda } \right)}}\;.
\label{e5.50}
\eeq
Here again, the spacetime geodesics, that is the $g_0$ family of paths, all have a $\hat f$ independent of operators, and using the normalization condition \rref{e5.49} above we trivially obtain the QGAP:
\beq
\left| \gamma \right|_q = \int_{}^\infty {dt\sqrt {\left\langle {\Psi \left| {\hat f} \right| \Psi } \right\rangle } } = \int_{}^\infty {dt\sqrt {\left\langle \Psi \; | \; \Psi \right\rangle f} } = \int_{}^\infty {dt\sqrt f }  < \infty \; ,
\label{e5.51}
\eeq
that is just the classical GAP value which is finite.

To calculate the  QGAP for all the other paths given above, we use the most general operator $\hat f$ function obtained in \rref{e5.40},
\beq
\hat f\left( {\hat \mu ,\hat \lambda , \hat a,\hat b; \;t} \right) \sim e^{ -2t} \left( {\cos\left[ {k\left( {e^{-t_0 } - e^{-t} } \right)\left( {\frac{1}{\hat a + \theta \hat b}} \right) \left( {\frac{\partial }{\partial b}-\theta \frac{\partial}{\partial a}} \right)} \right] + k^2 } \right) + O\left[ e^{-3t} \right].
\label{e5.52}
\eeq
Since for the general case we need to use the momentum space representation of operators, all we need to calculate the expectation values is to find the momentum representation of the modular invariant  wave functions for the torus universe. This is calculated below.

\subsubsection{Momentum Space Modular Invariant Wave Function}

We define position space and momentum space operators, based on the commutator \rref{e5.8}
\beq
\begin{array}{l}
 \hat {\!\vec{\,q}} = (\hat \mu , \hat \lambda ) \\ 
 \hat {\!\vec{\,P}} = ( - 2\hat a, 2\hat b) = -i\vec \nabla _{\vec q} \;, \quad \vec \nabla _{\vec q} = \left( {\frac{\partial}{\partial \mu }, \frac{\partial}{\partial \lambda}} \right) \\ 
\end{array}
\label{e5.53}
\eeq
and the complete set of position and momentum eigenstate basis $\left| {\mu ',\lambda '} \right\rangle$ and $\left| {a',b'} \right\rangle$ with completeness relations
\beq
1 = \int {d\mu \, {d\lambda \left| {\mu ,\lambda } \right\rangle \left\langle {\mu ,\lambda } \right|}},
\label{e5.54}
\eeq
\beq
1 = \int {d^2 p\left| {\vec p} \right\rangle } \left\langle {\vec p} \right| = - 4\int {da{\kern 1pt} {\kern 1pt} db} \left| {a,b} \right\rangle \left\langle {a,b} \right|.
\label{e5.55}
\eeq
Now by examining $\left\langle {\mu ',\lambda '} \right| \hat a \left| {a',b'} \right\rangle$ and $\left\langle {\mu ', \lambda '} \right| \hat b \left| {a',b'} \right\rangle$ we obtain a system of differential equations which can be solved easily
\beq
\left\{ {\begin{array}{*{20}l}
   {\hat a\left\langle {\mu ',\lambda '} \;|\; {a',b'} \right\rangle = a' \left\langle {\mu ', \lambda '} \; | \; {a',b'} \right\rangle = \frac{i}{2}\frac{\partial}{\partial \mu '} \left\langle {\mu ',\lambda '}\;|\; {a',b'} \right\rangle }  \\
   {\,\hat b\left\langle {\mu ',\lambda '} \;|\; {a',b'} \right\rangle = b' \left\langle {\mu ',\lambda '}\;|\; {a',b'} \right\rangle = - \frac{i}{2}\frac{\partial}{\partial \lambda '}\left\langle {\mu ',\lambda '} \;|\; {a',b'} \right\rangle }  \\
\end{array}} \right.
\label{e5.56}
\eeq
giving
\beq
\left\langle {\mu ',\lambda '} \;|\; {a',b'} \right\rangle = N\exp [2i\left( {\lambda ' b' - \mu ' a'} \right)]
\label{e5.57}
\eeq
with normalization constant $N = \frac{1}{2\pi}$, obtained by considering $\delta ^2 \left( \vec{q'} - \vec{q''} \right) = \left\langle \vec {q '} \; | \; \vec {q''} \right\rangle$. 

\noindent Thus
\beq
\Phi \left( {a,b} \right) = \left\langle {a,b} \; | \; {\Psi } \right\rangle = \frac{1}{2\pi} \int {d \mu {\kern 1pt} {\kern 1pt} d\lambda } \exp \left[ {2i\left( {\mu a - \lambda b} \right)} \right]\Psi \left( {\mu ,\lambda } \right)
\label{e5.58}
\eeq
where $\Psi \left( {\mu ,\lambda } \right) = \left\langle {\mu ,\lambda } \right| \left. \Psi \right\rangle$ is the position representation of the modular invariant wave function \rref{e5.44}. After some algebra we obtain
\beq
\Phi \left( {a,b} \right) = \int \frac{d^2 \tau}{{\tau _2} ^2} J\left( {T; \tau , \bar \tau ; a, b} \right) \, \bar{\tilde{\psi}} \left( {\tau ,\bar \tau ,T} \right)
\label{e5.59}
\eeq
where
\beq
J\left( {T; \tau , \bar \tau ; a, b} \right) = \frac{T \left( b - \tau a \right)}{2\pi {\kern 1pt} {\tau _2} ^{1/2} } \exp \left[ {i\frac{T}{\tau _2 } \left| {b - \tau a} \right|^2 } \right]
\label{e5.60}
\eeq
It can be easily verified that
\beq
\int {da\,db\,\bar J} \left( {T; \tau ', \bar \tau '; a, b} \right) J\left( {T; \tau , \bar \tau ; a, b} \right) = \frac{1}{4} {\tau _2}^2 \delta ^2 \left( {\tau - \tau '} \right),
\label{e5.61}
\eeq
and
\begin{eqnarray}
\left\langle \Phi \right| \hat f \left| \Phi \right\rangle & = & \int_{- \infty }^\infty {dP_1 \int_{-\infty}^\infty {dP_2 \; \bar \Phi \, \hat f \, \Phi}} \nonumber \\ 
   & = & 4\int_{-\infty}^\infty {da\int_{-\infty}^\infty {db \; \bar \Phi \left( {a,b} \right) \, \hat f \left( \hat a, \hat b \right) \, \Phi (a,b) }}\;. \label{e5.62}
\end{eqnarray}

\subsubsection{Calculation of the QGAP Using the Momentum Space Modular Invariant Wave Functions}

Now to find the QGAP for the paths $g_1$ to $g_4$ using \rref{e5.40}, we need to evaluate 
\begin{eqnarray}
&&\left| \gamma  \right|_q = \mathop {\lim }\limits_{t \to \infty } \int_{}^t {dt' \sqrt {\left\langle \Phi  \right| \hat f \left( {\hat \mu , \hat \lambda, \hat a, \hat b; t'} \right)\left| \Phi \right\rangle } } \nonumber \\ 
&& \sim \mathop {\lim }\limits_{t \to \infty } \int_{}^t \!{dt'e^{-t'} \sqrt {k^2 + 4 \mathop {\int}\limits_{\!\!\!-\infty }^{\;\infty} \!{da \mathop {\int}\limits_{\!\!\!-\infty }^{\;\infty} \!{db \; \bar \Phi \left( a, b \right) \, \cos\!\left[ {k \left( \frac{e^{-t_0}-e^{-t'}}{\hat a + \theta \hat b} \right)\!\!\left( \frac{\partial}{\partial b} - \theta \, \frac{\partial}{\partial a} \right)} \right]\Phi (a,b)}} }} \nonumber \\ 
&&\mbox{} + O\left[ {e^{-3t'/2} } \right]. \label{e5.63}
\end{eqnarray}
Expanding the cosine operator inside the integrals, near the singularity, we get
\begin{eqnarray}
&&\cos\left[ {k (e^{-t_0}-e^{-t'}) \left( \frac{1}{\hat a + \theta \hat b} \right) \left( \frac{\partial}{\partial b} - \theta \frac{\partial}{\partial a} \right)} \right] \simeq  \label{e5.64}\\ 
  &&\simeq 1 - \frac{1}{2!} k^2 e^{-2t_0} \left( \frac{1}{\hat a + \theta \hat b} \right)^2 \left( \frac{\partial}{\partial b} - \theta \frac{\partial }{\partial a} \right)^2 + \frac{1}{4!} k^4 e^{-4t_0} \left( \frac{1}{\hat a + \theta \hat b} \right)^4 \left( \frac{\partial}{\partial b} - \theta \frac{\partial}{\partial a} \right)^4  + ... \nonumber 
\end{eqnarray}
where we just kept the leading time dependent term in $ (e^{-t_0}-e^{-t'})^n=e^{-n t_0}(1-e^{t'-t_0})\simeq e^{-n t_0}$ which is a valid approximation near the singularity. Thus we obtain
\begin{eqnarray}
&&\left| \gamma \right|_q \sim \nonumber\\
&&\mathop {\lim }\limits_{t \to \infty } \int_{}^t {\!\! dt'e^{-t'} \sqrt {1 \!+ \!k^2 \! - \! 2k^2 \! e^{-2t_0} \!\mathop {\int}\limits_{\!\!\!-\infty }^{\;\infty} {\! da\mathop {\int}\limits_{\!\!\!-\infty }^{\;\infty} {\! db \; \bar \Phi \left( a,b \right) \left( \frac{1}{a + \theta b} \right)^{\!2} \! \left( \frac{\partial}{\partial b} - \theta \frac{\partial}{\partial a} \right)^{\! 2} \! \Phi (a,b)}} + O\left[ e^{-4t_0} \right]} }  \nonumber\\ 
&& \mbox{} + O\left[ e^{-3t'/2} \right] \nonumber\\ 
&&\sim \mathop {\lim }\limits_{t \to \infty } \sqrt {1 \! + \! k^2 }\! \int_{}^t \! dt' e^{-t'} \! \left( 1 \! - \! \frac{k}{\sqrt {1\! + \!k^2 } } e^{-t_0 }\! \mathop {\int}\limits_{\!\!\!-\infty }^{\;\infty}\! da \! \mathop {\int}\limits_{\!\!\!-\infty }^{\;\infty}\! db \;\bar \Phi \, \left( \frac{1}{a + \theta b} \right)^{\! 2} \left( \frac{\partial}{\partial b}-\theta \frac{\partial}{\partial a} \right)^{\! 2} \!\Phi \right. \nonumber \\
&& \mbox{} + \! O\left[ e^{-3t'/2} , e^{-3t_0 /2} \right] \Bigg). \label{e5.65}
\end{eqnarray}
Here, we can choose our starting point at $t_0$ close enough to the singularity that $e^{-t_0}$ is sufficiently small and we ignore it in the first approximation to the QGAP. We see that to the first order we obtained our old value
\beq
\left| \gamma \right|_q \mathop {\sim \lim }\limits_{t \to \infty } \sqrt {1 + k^2 } \int_{}^t {dt'e^{- t'} } < \infty
\label{e5.66}
\eeq
which is certainly finite, implying that the QGAP is finite if we start sufficiently (but finitely) close to the singularity. More interestingly, up to this order we see clearly the independence of the QGAP from the particular wave function. We would expect that the higher order corrections that depend on the wave functions would only contribute to the specific finite value of the QGAP, which is irrelevant, but will not make it divergent. To verify this we need to examine the higher terms more carefully. However, since the wave function here is time dependent and rather complicated, it is not easy to evaluate the first few, let alone, all the terms. Besides, exact evaluation of the first few terms would not obviously be sufficient since the finiteness of QGAP would depend on the convergence of all the terms. 

However, we can make some general observations about all the terms, without evaluating them exactly. 

First, note that the $a$ and $b$ dependence inside the double integrals are not of interest here since their contribution is exactly the same when we consider the QGAP for a finite path, far from the singularity with endpoints $t_1$ and $t_2$, where we do not expect the overall QGAP to diverge. The important contribution is due to the time-dependent behavior of the integral, which is sensitive to approach to the singularity. 

Now we consider the time dependent factors. Obviously, the $e^{-t_0}-e^{-t} \approx e^{-t_0}$ term in \rref{e5.63} should not cause a problem. This is because of the following simple argument. For a finite path between $t_1$ and $t_2$ (with $t_2 > t_1$), which is far from the singularity and hence has a finite QGAP, this factor is of form:  $e^{-t_1}-e^{-t_2}$. We can always choose a path approaching the singularity at $t \to \infty$, starting from an initial point $t_0 \geq t_1 + \ln{(1-e^{-(t_2-t_1)})^{-1}}> t_1$. With this choice, we always have $e^{-t_0}-e^{-t}\leq e^{-t_1}-e^{-t_2}$, if we start sufficiently close to the singularity. So, we would not expect this factor by itself to cause the QGAP to diverge.

We should also look at the time-dependence, when the derivatives in \rref{e5.64} act on the momentum space modular invariant wave function \rref{e5.49}. In this case, the time-dependant effects can be only due to $J\left( {T; \tau , \bar \tau ; a, b} \right)$ given by \rref{e5.60}, and $\tilde{\psi} \left( {\tau , \bar \tau , T} \right)$ given by \rref{e5.46}. We can see that in both of these time-dependent pieces, there are oscillatory time-dependent exponentials, that act destructively as we approach the singularity as $t \to \infty$. Now, the derivatives are all with respect to $a$ and $b$, and therefore, can act only on $J\left( {T; \tau , \bar \tau ; a, b} \right)$ piece. Each derivative can, at most, bring down a factor of $T$, resulting in terms with $T$ to some power, multiplied by an oscillatory exponential factor, with a $T$ in the exponent. Recall that $T\approx e^t$ near the singularity through \rref{e5.5}. Therefore, we would expect as we approach the singularity at $t \to \infty$ ($T \to \infty$), that the destructive interference of $\exp \left[ {i\frac{T}{\tau _2 } \left| {b - \tau a} \right|^2 } \right]$ factor should cancel the effects of $T^n$ factors. The only other time-dependence is outside the square root in the form of $e^{-t}$ which will suppress the integrand further as we approach the singularity.

Therefore, with reasonable approximations, we can make statements about all the terms in QGAP in this case. These arguments suggest that the QGAP should stay finite as we approach the singularity, and as before, all families of paths, $g_0$ to $g_4$, remain b-incomplete. The singularity seems to persist for all the paths, also for the modular invariant wave function. 

An exact evaluation with consideration of convergence would be desirable to make a more conclusive statement about this wave function. Some work in this direction is in progress.

\section{Caveats}

In this approach we used a particular \textit{definition} to examine the issue of singularities. Although Hosoya's definition of the QGAP is a reasonable generalization of the GAP which is widely used to examine existence of singularities, it is not a unique generalization, and b-completeness is not the only possible criteria.

Many \textit{paths} have been examined here, but in no way is this exhaustive. In classical general relativity it is sufficient to have only one causal path that is incomplete to indicate a singularity. Although it seems that at least in the case of the torus universe, a very large class of family of paths are incomplete, this is still only suggestive. In quantum theory, a test particle travels through all paths, in a path integral sense; large classes of complete paths may only contribute a measure zero, although this does not seem to be likely for the torus universe.
     
The QGAP, and possibly many conceivable quantum criteria for singularities, are determined in terms of the \textit{wave function}, and persistence or resolution of singularities can depend on the state. But with the different choices of states used here, the overall behavior seemed to be independent of the particular states.
    
Although two different \textit{spacetime models} with quite different singular characters were examined here, it may be that other (2+1)-dimensional models would resolve the singularities.

We used (2+1)-dimensional canonical quantum gravity here to study the singularities. There are many different \textit{theories of quantum gravity} that are not equivalent, and there are many \textit{schemes} within each approach. There is no guarantee that conclusions about singularities will be consistent in all these approaches, and there is no indication yet as which approach is more realistic.

The general characteristics and limitations of \textit{the (2+1) dimensional spacetime} itself may be the critical factor responsible for persistence of singularities. It may be that the richer structure of the realistic (3+1)-dimensional spacetime could result in resolution of singularities.

\section{Comments}

It should be mentioned that persistence of spacetime singularities is not totally unexpected or even undesirable. Penrose has argued that even in a quantum theory of gravity, the big bang singularity should persist in some form \cite{example}. Horowitz and Myers, have also argued \cite{horowitz&myers} that in quantum gravity, in order to have a stable ground state, some singularities (such as negative mass black hole solutions) must not be resolved. In general they argue that any field theory will contain some curvature singularities, and that certain classes of timelike singularities can not be resolved in a physically reasonable theory.

\section{Conclusions}

Previous studies of spacetime singularities have looked at the behavior of quantum test particles and matter fields probing the singularities in classical spacetime background. This work focuses on the spacetime itself when it becomes quantized. The most troubling aspect about the spacetime singularities is that the description of spacetime breaks down in classical regime. The QGAP is an appropriate tool to test if fluctuations of spacetime near a classically singular region can smear out the singular behavior.

I have shown that such fluctuations do not seem to smear out the singularities in two very different models of (2+1)-dimensional spacetime. This was demonstrated using many different families of paths, timelike and lightlike, geodesic and accelerated, with varying speed parameters. The different wave functions considered here did not seem to alter the singular nature of the geometry. This seems to give credence to the idea that the resolution or persistence of singularities in quantum gravity may not depend on the particular quantum state (which could in principle be manipulated), but may be of a more permanent nature, determined perhaps only by the particular quantum gravity theory.

An interesting feature, and a cautionary note, is that at least for some milder singularities in the quantum regime, some paths that were classically incomplete become complete, as Hosoya has shown. However, further analysis suggested that  this did not seem to be the predominant character. This may explain the seemingly conflicting results obtained about resolution of singularities in different methods and models. It may also be that spacetime fluctuations smear out some mild singularities, but not others. Such a compromising and democratic possibility may even appeal to both opponents and proponents of singularities.

\section{Future Directions}

The caveats given above suggests direction for future work in this area. The definition used is not unique. Many different approaches, perhaps with other quantum generalizations of classical criteria, can be tried. Other singular spacetimes need to be studied to develop a comprehensive characterization of singularities within the context of quantum gravity. This may reveal a richer and varied nature of quantum behavior of singularities. The various existing inequivalent quantization approaches may each pass a different verdict on the question of singularities. This in itself might give us a criterion to judge competing theories of quantum gravity, or even serve as the basis for experimental pruning and verification. More elaborate approaches in 2+1 gravity with matter fields may also result in a more realistic assessment.

\section*{Acknowledgements}
I would like to thank Steven Carlip for many useful discussions, for carefully reading drafts of this paper, and for his valuable suggestions and comments. This work was supported in part by US Department of Energy grant DE-FG03-91ER40674.

\end{document}